\documentclass[final,1p,times]{elsarticle}

 \usepackage{graphicx}
 \usepackage{color}

\usepackage{amssymb}

\usepackage{lineno}
\usepackage{bm}

\journal{Solid State Communications}

\begin{document}

\def\gsim{\mathop {\vtop {\ialign {##\crcr 
$\hfil \displaystyle {>}\hfil $\crcr \noalign {\kern1pt \nointerlineskip } 
$\,\sim$ \crcr \noalign {\kern1pt}}}}\limits}
\def\lsim{\mathop {\vtop {\ialign {##\crcr 
$\hfil \displaystyle {<}\hfil $\crcr \noalign {\kern1pt \nointerlineskip } 
$\,\,\sim$ \crcr \noalign {\kern1pt}}}}\limits}

\begin{frontmatter}



\title{Non-Divergent Gr\"{u}neisen Parameter in Quantum Critical Quasicrystal Yb$_{15}$Al$_{34}$Au$_{51}$:
Reflection of Robustness of Quantum Criticality under Pressure}


\author{Shinji Watanabe and Kazumasa Miyake$^2$}

\address{Department of Basic Sciences, Kyushu Institute of Technology, Kitakyushu, Fukuoka  804-8550, Japan \\
$^2$Center for Advanced High Magnetic Field Science, Osaka University, Toyonaka, Osaka 560-0043, Japan}

\begin{abstract}
The mechanism of {\it not} diverging Gr\"{u}neisen parameter in the quantum critical heavy-fermion quasicrystal (QC) Yb$_{15}$Al$_{34}$Au$_{51}$ is analyzed. 
We construct the formalism for calculating the specific heat $C_V(T)$, the thermal-expansion coefficient $\alpha(T)$, and the Gr\"{u}neisen parameter $\Gamma(T)$ near the quantum critical point of the Yb valence transition. 
By applying the framework to the QC, we calculate $C_V(T)$, $\alpha(T)$, and $\Gamma(T)$, which explains the measurements. 
{\it Not} diverging $\Gamma(T)$ is attributed to the robustness of the quantum criticality in the QC under pressure. 
The difference in $\Gamma(T)$ at the lowest temperature between the QC and approximant crystal is shown to reflect the difference in the volume derivative of characteristic energy scales of the critical Yb-valence fluctuation and the Kondo temperature. 
Possible implications of our theory to future experiments are also discussed. 
\end{abstract}


\end{frontmatter}



\section{Introduction}
Quantum critical phenomena have been one of the central issues in condensed matter physics. 
Recent discovery of unconventional quantum criticality in heavy-fermion quasicrystal (QC) Yb$_{15}$Al$_{34}$Au$_{51}$ has attracted great interest~\cite{Deguchi,Watanuki}. The measured criticality of the magnetic susceptibility $\chi\sim T^{-0.5}$, the specific heat $C/T\sim-\ln{T}$, the resistivity $\rho\sim T$, and the NMR relaxation rate $(T_{1}T)^{-1}\sim T^{-0.5}$ 
is similar to those observed in periodic crystals such as YbCu$_{5-x}$Al$_x$ $(x=1.5)$~\cite{Bauer}, YbRh$_2$Si$_2$~\cite{Trovarelli}, $\beta$-YbAlB$_4$~\cite{Nakatsuji}, and $\alpha$-YbAl$_{1-x}$Fe$_{x}$B$_4$ $(x=0.014)$~\cite{Kuga2018}, which have been proposed to be explained by the theory of critical valence fluctuations (CVF) of Yb~\cite{WM2010}.  

\begin{figure}
\includegraphics[width=10cm]{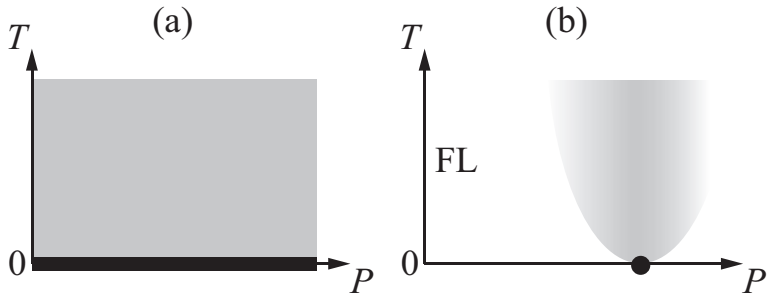}
\caption{(color online) 
Schematic $T$-$P$ phase diagrams of (a) QC Yb$_{15}$Al$_{34}$Au$_{51}$ and (b) AC Yb$_{14}$Al$_{35}$Au$_{51}$. Thick solid line in (a) represents condensation of the QCPs. In (b), closed circle indicates the QCP and FL indicates Fermi-liquid region. Shaded areas represent quantum critical regions. 
}
\label{fig:T_P}
\end{figure}

The quantum criticality in the QC Yb$_{15}$Al$_{34}$Au$_{51}$ emerges without tuning 
and surprisingly persists even under pressure at least up to $P=1.6$~GPa [Fig.~\ref{fig:T_P}(a)]~\cite{Deguchi}. 
The theory of CVF for the QC has shown that the valence quantum critical points (QCPs) are condensed forming a cluster 
in the ground-state phase diagram, which explains the criticality robust against pressure~\cite{WM2013}. 


Interestingly, zero-tuning criticality and its robustness under pressure were also  reported in $\beta$-YbAlB$_4$~\cite{Nakatsuji,Tomita2015}. Recently, the direct evidence of the valence QCP has been observed in the sister compound $\alpha$-YbAl$_{1-x}$Fe$_{x}$B$_4$ $(x=0.014)$~\cite{Kuga2018}. 

As for the 
general 
property of the QCP, 
the Gr\"{u}neisen parameter~\cite{Gruneisen} 
\begin{eqnarray}
\Gamma=\frac{\alpha V}{C_{V}\kappa_T},
\label{eq:Grn}
\end{eqnarray}
where $C_V$ is the specific heat under a constant volume $V$, $\alpha$ is the thermal-expansion coefficient, and $\kappa_T$ is the isothermal compressibility, 
has been used widely as a clear indicator of whether the material is located at the QCP 
since it was asserted that the Gr\"{u}neisen parameter diverges at {\it any} QCPs by the renormalization-group theory~\cite{Zhu2003,Garst}. 
Divergence of the Gr\"{u}neisen parameter at the QCP was observed in several materials~\cite{Gegenwart2016} such as CeNi$_2$Ge$_2$~\cite{Kuchler2003}, YbRh$_2$(Si$_{0.95}$Ge$_{0.05}$)$_2$~\cite{Kuchler2003}, CeCu$_{5.8}$Ag$_{0.2}$~\cite{Kuchler2004}, and CeCu$_{5.9}$Au$_{0.1}$~\cite{Grube2017}. 
However, recent measurement in the QC Yb$_{15}$Al$_{34}$Au$_{51}$ has revealed that $|\Gamma|$ does not show such a divergence as lowering the temperature but rather decreases and finally converges  into a finite value $\Gamma\approx -55$ at the lowest temperature $T=70$~mK~\cite{MGS}. 
This poses a serious challenge to theories ever proposed. 
Furthermore, the Gr\"{u}neisen parameter in the 1/1 approximant crystal (AC) Yb$_{14}$Al$_{35}$Au$_{51}$ 
with the periodic lattice structure
showing the Fermi-liquid behavior 
has been observed as $\Gamma\approx -130$ at $T=70$~mK, whose absolute value is much larger than that in the QC~\cite{MGS}. 

In this paper, we present an explanation for resolving these puzzles from the viewpoint of the theory of CVF. The CVF theory is consistent with  
the criticality in the magnetic susceptibility and its robustness under pressure in the QC~\cite{WM2013} 
as well as emergence of the same criticality in the pressure-tuned AC 
[Fig.~\ref{fig:T_P}(b)]
\cite{WM2016}. 
First, we will construct a general formalism for calculating $C_V$, $\alpha$, and $\Gamma$ in the  thermodynamically consistent way near the valence QCP in the periodic lattice systems in Sect.~2. 
Next, by applying the framework to the QC, 
we will discuss the properties of $C_V$, $\alpha$, and $\Gamma$ in the QC in Sect.~3. 
In Sect.~4, we will discuss the physical meaning of the Gr{\"u}neisen paramters observed in the QC and AC at ambient pressure. In Sect.~5, we will present theoretical predictions for future experiments. 
The paper will be summarized in Sect.~6. 

\section{Formulation near the valence QCP}
Recently, the complete expressions of the thermal expansion $\alpha$ and the Gr{\"u}neisen parameter $\Gamma$ near the magnetic QCP have been derived on the basis of the self-consistent renormalization (SCR) theory for spin fluctuations~\cite{WM2018SCR,WM2019SCR}. 
In this section, 
by extending the formalism for spin fluctuations we recapitulate the formulation for calculating $\alpha$ and $\Gamma$ near the valence QCP in the three spatial dimension. 
Below the energy units are taken as $\hbar=1$ and $k_{\rm B}=1$ unless otherwise noted. 

The valence transition and the QCP are described in the extended periodic Anderson model, which is defined by the periodic Anderson model with the Coulomb repulsion between f and conduction electrons~\cite{WM2010}. 
By applying the Stratonovich-Hubbard transformation to the inter-orbital Coulomb-repulsion term, the following action 
\begin{eqnarray}
\Phi\left[\varphi\right]&=&\sum_{m}\left[
\frac{1}{2}
\sum_{\bar{q}}\Omega_{2}(\bar{q})
\varphi_{m}(\bar{q})\varphi_{m}(-\bar{q})
\right.
\nonumber
\\
& &
+
\left.
\sum_{\bar{q}_1,\bar{q}_2,\bar{q}_3}
\Omega_{3}(\bar{q}_1,\bar{q}_2,\bar{q}_3)
\varphi_{m}(\bar{q}_1)
\varphi_{m}(\bar{q}_2)
\varphi_{m}(\bar{q}_3)
\delta\left(\sum_{i=1}^{3}\bar{q}_{i}\right)
\right.
\nonumber
\\
& &+ 
\left.
\sum_{\bar{q}_1,\bar{q}_2,\bar{q}_3,\bar{q}_4}
\Omega_{4}(\bar{q}_1,\bar{q}_2,\bar{q}_3,\bar{q}_4)
\varphi_{m}(\bar{q}_1)
\varphi_{m}(\bar{q}_2)
\varphi_{m}(\bar{q}_3)
\varphi_{m}(\bar{q}_4)
\delta\left(\sum_{i=1}^{4}\bar{q}_{i}\right)
\right.
\nonumber
\\
& &
+\left.\cdots
\right] 
\label{eq:Action}
\end{eqnarray}
is obtained 
in the notation $\bar{q}\equiv({\bm q},i\omega_{l})$ with $\omega_{l}=2l\pi T$ $(l=0, \pm{1}, \pm{2}, \cdots)$~\cite{WM2010}. 
Starting from Eq.~(\ref{eq:Action}), we constructed the 
SCR 
theory for CVF 
in periodic lattice systems~\cite{WM2010}. 
Since the long-wave-length $|{\bm q}|\ll q_{\rm c}$ around ${\bm q}={\bf 0}$ 
and low-frequency $|{\omega}|\ll \omega_{\rm c}$ region plays the dominant role  
in the critical phenomena with $q_{\rm c}$ and $\omega_{\rm c}$ being cutoffs 
for the momentum and frequency in the order of inverse of the lattice constant 
and the effective Fermi energy, respectively, 
$\Omega_{i}$ for $i=2,3$, and 4 are expanded for $q$ and $\omega$ around 
$({\bf 0},0)$: 
\begin{eqnarray}
\Omega_{2}({\bm q},i\omega_{l})
\approx 
\eta_0
+Aq^{2}+C_q\left|\omega_{l}\right|, 
\label{eq:O2expand}
\end{eqnarray}
with $C_q\equiv C/q$, 
$\Omega_{3}(\bar{q}_1,\bar{q}_2,\bar{q}_3)\approx v_3/\sqrt{\beta N}$, and 
$\Omega_{4}(\bar{q}_1,\bar{q}_2,\bar{q}_3,\bar{q}_4)\approx v_4/(\beta N)$, 
where $N$ is the number of Yb atoms per mole and $\beta$ is defined as $\beta\equiv 1/T$.
Here, we consider the isotropic system.   
Then, taking account of the mode-mode coupling effects up to the 4th order in $\Phi[\varphi]$ 
in Eq.~(\ref{eq:Action}), 
we employ Feynman's inequality for the free energy: 
\begin{eqnarray}
F\le F_{\rm eff}+T\langle \Phi-\Phi_{\rm eff}\rangle_{\rm eff}\equiv\tilde{F}(\eta), 
\label{eq:Free_SCR}
\end{eqnarray}
where 
$\Phi_{\rm eff}$ is the effective action for the best Gaussian, 
\begin{eqnarray}
\Phi_{\rm eff}[\varphi]=\frac{1}{2}\sum_{m}\sum_{{\bf q},l}
\chi_{{\rm v}}({\bm q},{i}\omega_{l})^{-1}
|\varphi_{m}({\bm q},{i}\omega_{l})|^2.
\label{eq:Phi_eff}
\end{eqnarray}
In Eq.~(\ref{eq:Free_SCR}),   
$\langle\cdots\rangle_{\rm eff}$ denotes the statistical average taken by the weight $\exp\left(-\Phi_{\rm eff}[\varphi]\right)$ and $F_{\rm eff}$ is given by  
\begin{eqnarray}
F_{\rm eff}=-T\ln\int{\cal D}\varphi\exp\left(-\Phi_{\rm eff}[\varphi]\right). 
\label{eq:F_eff}
\end{eqnarray}
In Eq.~(\ref{eq:Phi_eff}), 
the valence susceptibility $\chi_{{\rm v}}({\bm q},{i}\omega_{l})$ is defined as 
\begin{eqnarray}
\chi_{\rm v}({\bm q},{i}\omega_{l})^{-1}
=\eta+Aq^2+C_{q}|\omega_{l}|, 
\label{eq:chi_v}
\end{eqnarray}
%
where $\eta$ expresses the effect of the
 mode-mode coupling of CVF and parametrizes the closeness to the QCP. 
The free energy $\tilde{F}$ defined by Eq.~(\ref{eq:Free_SCR}) is expressed as
\begin{eqnarray}
\tilde{F}&=&\frac{1}{\pi}\sum_{q}\int_{0}^{\omega_{\rm c}}d\omega
\frac{\Gamma_{q}}{\omega^2+\Gamma_{q}^2}\left\{\frac{\omega}{2}+T{\ln}\left(1-{\rm e}^{-\frac{\omega}{T}}\right)\right\}
\nonumber
\\
&+&\frac{\eta_0-\eta}{2}\langle\varphi^2\rangle_{\rm eff}
+\frac{3v_{4}}{N}\langle\varphi^2\rangle_{\rm eff}^2
-\frac{1}{\pi}\sum_{q}\frac{\pi\omega_{\rm c}}{4},     
\label{eq:freeE}
\end{eqnarray}
where $\Gamma_{q}$ is defined by $\Gamma_{q}\equiv(\eta+Aq^2)/C_q$ and valence fluctuation $\langle\varphi^2\rangle_{\rm eff}$ is defined as
\begin{eqnarray}
\langle\varphi^2\rangle_{\rm eff}=T\sum_{q}\sum_{l}\frac{1}{\eta+Aq^2+C_q|\omega_{l}|}.
\label{eq:phi2_def}  
\end{eqnarray}
Here, $\langle\varphi^2\rangle_{\rm eff}$ 
consists of the quantum (zero-point) fluctuation $\langle\varphi^2\rangle_{\rm zero}$ and thermal fluctuation $\langle\varphi^2\rangle_{\rm th}$ as 
\begin{eqnarray}
\langle\varphi^2\rangle_{\rm eff}=
\langle\varphi^2\rangle_{\rm zero}+\langle\varphi^2\rangle_{\rm th},  
\label{eq:psi2}
\end{eqnarray}
where $\langle\varphi^2\rangle_{\rm zero}$ and $\langle\varphi^2\rangle_{\rm th}$ are expressed as 
\begin{eqnarray}
\langle\varphi^2\rangle_{\rm zero}&=&
\frac{1}{\pi}\sum_{q}\frac{1}{C_q}\int_{0}^{\omega_{\rm c}}d\omega
\frac{\omega}{\Gamma_{q}^2+\omega^2}, 
\label{eq:phi2_zero}
\\
\langle\varphi^2\rangle_{\rm th}&=&
\frac{1}{\pi}\sum_{q}\frac{2}{C_q}\int_{0}^{\omega_{\rm c}}d\omega
\frac{1}{{\rm e}^{\beta\omega}-1}
\frac{\omega}{\Gamma_{q}^2+\omega^2}, 
\label{eq:phi2_th}
\end{eqnarray}
respectively. 

Near the QCP of the valence transition, quantum valence fluctuation   
$\langle\varphi^2\rangle_{\rm zero}$  in Eq.~(\ref{eq:phi2_zero}) is calculated as 
\begin{eqnarray}
\langle\varphi^2\rangle_{\rm zero}=N\frac{3T_0}{2T_A}
C_1-C_{2}y+\cdots,
\label{eq:phi2_zero_b}
\end{eqnarray}
where the characteristic temperature of valence fluctuation is defined as
\begin{eqnarray}
T_0\equiv\frac{Aq_{\rm B}^2}{2\pi C_{q_{\rm B}}}
\label{eq:T0}
\end{eqnarray}
%
with $q_{\rm B}$ being the wave number of the Brillouin Zone.  
In Eq.~(\ref{eq:phi2_zero_b}), $T_{A}$ is defined as
\begin{eqnarray}
T_{A}\equiv\frac{Aq_{\rm B}^2}{2}.      
\label{eq:TA}
\end{eqnarray}
and $y$ is defined as
\begin{eqnarray}
y\equiv \frac{\eta}{Aq_{\rm B}^2}.
\label{eq:y_def}
\end{eqnarray}
The constants $C_1$ and $C_2$ in Eq.~(\ref{eq:phi2_zero_b}) are given by
\begin{eqnarray}
C_1&=&
\int_{0}^{x_{\rm c}}dxx^{3}\ln\left|\frac{\omega_{{\rm c}T_0}^2+x^{6}}{x^{6}}\right|,
\label{eq:C1}
\\
C_2&=&
2\int_{0}^{x_{\rm c}}dxx\frac{\omega_{{\rm c}T_0}^2}{\omega_{{\rm c}T_0}^2+x^{6}}, 
\label{eq:C2}
\end{eqnarray}
respectively, where $x$ is the dimensionless wave number defined as $x\equiv q/q_{\rm B}$. 
Here,    
the cut off is expressed as $x_{\rm c}\equiv q_{\rm c}/q_{\rm B}$ in the dimensionless scaled form and 
$\omega_{{\rm c}T}$ is defined as $\omega_{{\rm c}T}\equiv\omega_{\rm c}/(2\pi{T})$.

The thermal valence fluctuation   
$\langle\varphi^2\rangle_{\rm th}$ in Eq.~(\ref{eq:phi2_th}) is calculated as
\begin{eqnarray}
\langle\varphi^2\rangle_{\rm th}=3N\frac{T_0}{T_A}
\int_{0}^{x_{\rm c}}dxx^{3}
\left\{
{\ln}u-\frac{1}{2u}-\psi(u)
\right\},  
\label{eq:psi2_th}
\end{eqnarray}
where $\psi(u)$ is the digamma function 
with 
$u$ being defined as 
\begin{eqnarray}
u\equiv\frac{\Gamma_{q}}{2\pi{T}}=\frac{x(y+x^2)}{t}. 
\label{eq:u}
\end{eqnarray}
Here, $t$ is defined as the dimensionless scaled temperature 
\begin{eqnarray}
t\equiv\frac{T}{T_{0}}. 
\label{eq:t}
\end{eqnarray}
%

Under the optimal condition $\frac{d\tilde{F}(\eta)}{d\eta}=0$, the SCR equation in the $Aq_{\rm B}^2\lsim\eta$ regime with $q_{\rm B}$ being the wavenumber of the Brillouin zone is obtained as 
\begin{eqnarray}
y=y_{0}+\frac{3}{2}y_{1}t\left[
\frac{x_{\rm c}^3}{6y}
-\frac{1}{2y}\int_{0}^{x_{\rm c}}dx\frac{x^3}{x+\frac{t}{6y}}
\right], 
\label{eq:SCReq}
\end{eqnarray}
where $y_{0}$ and $y_{1}$ are dimensionless constants~\cite{WM2010}. 
On the basis of the extended periodic Anderson model, 
it was shown that the almost dispersionless CVF mode appears around ${\bm q}={\bf 0}$ because of the strong local correlation effect for f electrons so that extremely small $A$ is realized in Eq.~(\ref{eq:chi_v})~\cite{WM2010}. This local character of the CVF is reflected in the valence SCR equation [Eq.~(\ref{eq:SCReq})] whose solution shows the new type of quantum criticality.
At the valence QCP, $y_0=0$ is realized in Eq.~(\ref{eq:SCReq}), where the valence susceptibility 
$\chi_v({\bf 0},0)\propto y^{-1}$  
diverges toward $t=0$, implying the diverging CVF.

It is noted that the valence susceptibility can have a maximum at finite $\bm Q$. In that case, $C_q$ in Eq.~(\ref{eq:chi_v}) is given by $C_q=C$~\cite{WM2010}. Even in that case almost flat dispersion of the CVF appears around ${\bm q}={\bm Q}$ because of strong local correlation for f electrons, giving rise to the extremely small $A$. The localness of the CVF still  yields the new type of the quantum criticality. Below we will show the formulation for $C_q=C/q$.

The entropy $S=-\left(\frac{\partial\tilde{F}}{\partial T}\right)_V$ is obtained from the free energy $\tilde{F}(y)$ in Eq.~(\ref{eq:freeE}) with the stationary condition of the SCR theory being satisfied, 
which results in 
\begin{eqnarray}
S&=&-3N\int_{0}^{x_{\rm c}}dxx^{2}
\left\{
{\ln}\sqrt{2\pi}-u+\left(u-\frac{1}{2}\right){\ln}u-{\ln}\Gamma(u)
\right\}
\nonumber
\\
&+&3N\int_{0}^{x_{\rm c}}dxx^{2}u
\left\{
{\ln}u-\frac{1}{2u}-\psi(u)
\right\}.
\label{eq:S}
\end{eqnarray}
Here, $\Gamma(u)$ is the Gamma function.  

Then, the 
specific heat is derived from $C_V=T\left(\frac{\partial S}{\partial T}\right)_V$ as 
\begin{eqnarray}
C_V=C_{\rm a}-C_{\rm b},
\label{eq:Cv}
\end{eqnarray}
where 
$C_{\rm a}$ and $C_{\rm b}$ are given by 
%
\begin{eqnarray}
C_{\rm a}&=&-3N\int_{0}^{x_{\rm c}}dxx^{2}u^2\left\{
\frac{1}{u}+\frac{1}{2u^2}-\psi'(u)
\right\}, 
\label{eq:Ca}
\\
C_{\rm b}&=&
\tilde{C}_{\rm b}
\left(\frac{\partial{y}}{\partial{t}}\right)_{V},
\label{eq:Cb}
\end{eqnarray}
%
respectively~\cite{Takahashi}.  
In Eq.~(\ref{eq:Ca}),  
$\psi'(u)$ is the trigamma function.
The explicit form of $\tilde{C}_{\rm b}$ is given by 
%
\begin{eqnarray}
\tilde{C}_{\rm b}=-3N\int_{0}^{x_{\rm c}}dxx^{3}u\left\{
\frac{1}{u}+\frac{1}{2u^2}-\psi'(u)
\right\}. 
\label{eq:tildaCb}
\end{eqnarray}
%
  
The explicit form of  
the temperature dependent factor $\left(\frac{\partial y}{\partial t}\right)_V$ is obtained by differentiating Eq.~(\ref{eq:SCReq}) with respect to $t$ under a constant volume as
%
\begin{eqnarray}
\left(
\frac{\partial y}{\partial t}\right)_V=
\frac{3}{2}y_1
\frac{
\frac{I_1}{y}+\frac{tI_2}{12y^2} 
}
{
1+\frac{3}{2}y_{1}t\left(\frac{I_1}{y^2}+\frac{tI_2}{12y^3}\right)
},
\label{eq:dydT}
\end{eqnarray}
%
where $I_1$ is 
given by $I_1=\frac{x_{\rm c}^3}{6y}-\frac{1}{2y}\int_{0}^{x_{\rm c}}dx\frac{x^3}{x+\frac{t}{6y}}$ 
and $I_2$ is given by $I_2=\int_{0}^{x_{\rm c}}dx\frac{x^3}{\left(x+\frac{t}{6y}\right)^2}$.

The thermal-expansion coefficient $\alpha$ is expressed as    $\alpha=-\frac{1}{V}\left(\frac{\partial{S}}{\partial{P}}\right)_{T}$ through the Maxwell relation $\left(\frac{\partial V}{\partial T}\right)_P=-\left(\frac{\partial S}{\partial P}\right)_T$, and is given in the form  
$\alpha=\alpha_{ a}+\alpha_{ b}$, 
where $\alpha_{ a}$ and $\alpha_{ b}$ are given by 
\begin{eqnarray}
\alpha_{\rm a}&=&\frac{1}{V}\frac{C_{\rm a}}{T_0}\left(\frac{\partial{T_0}}{\partial{P}}\right)_{T}
\label{eq:a_a},
\\
\alpha_{\rm b}&=&\frac{1}{V}\frac{\tilde{C}_{\rm b}}{t}\left(\frac{\partial{y}}{\partial{P}}\right)_{T},
\label{eq:a_b}
\end{eqnarray}
respectively, with $V$ being the molar volume. 

The Gr\"{u}neisen parameter is obtained by substituting Eq.~(\ref{eq:Cv}) and the above expression for $\alpha$ into Eq.~(\ref{eq:Grn}). 

\section{Application to quasicrystal}
\begin{figure}
\includegraphics[width=12cm]{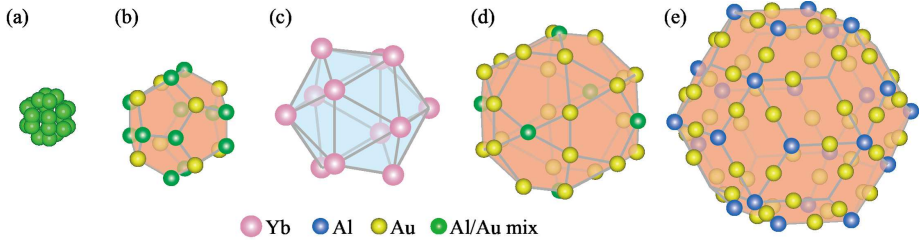}
\caption{(color online) Concentric shell structures of Tsai-type cluster in the Yb-Al-Au approximant~\cite{Deguchi,Ishimasa}:
(a) first shell, (b) second shell, (c) third shell, (d) fourth shell, and (e) fifth shell.  
}
\label{fig:Tsai_cluster}
\end{figure}

We now discuss the QC Yb$_{15}$Al$_{34}$Au$_{51}$. 
The QC and AC consist of the Tsai-type cluster, which has concentric shell structures from the 1st to 5th shells, as shown in Fig.~\ref{fig:Tsai_cluster}~\cite{Ishimasa,Deguchi}. 
The 
1/1
AC has the periodic arrangement of the Tsai-type cluster with the body-centered cubic (bcc) structure. 
In the $F_{n-1}/F_{n-2}$ AC [$F_n$ is the Fibonacci number i.e., $F_{n}$=$F_{n-1}$+$F_{n-2}$ ($F_1=1$, $F_2=1$, $n\ge 3$)], as $n$ increases, the size of the unit cell increases and the $n\to\infty$ limit corresponds to the QC~\cite{Goldman1993}.

The theoretical analysis of the periodic Anderson model in the AC has shown that the charge-transfer 
(CT)
fluctuation between the 4f electron at Yb on the icosahedron (3rd shell) 
[see Fig.~\ref{fig:Tsai_cluster}(c)]
 and 3p electron at Al on just the outer icosidodecahedron (4th shell) 
[see Fig.~\ref{fig:Tsai_cluster}(d)]
 is considerably enhanced~\cite{WM2016}. 
The mode-mode coupling theory applied to the most dominant 
CT 
fluctuation, i.e., CVF, has shown that the magnetic susceptibility
%
\begin{eqnarray}
\chi(T,B=0)
\propto\chi_{\rm v}(T,B=0)\propto y
\sim T^{-0.5}
\end{eqnarray}
%
and the $T/B$ scaling, where $\chi(T,B)$ is expressed as a single scaling function of the ratio of temperature $T$ and magnetic field $B$, appear at the pressure-tuned valence QCP~\cite{WM2016}. 
This theoretical prediction~\cite{WM2016} has actually been confirmed in the AC at $P=1.96$~GPa
[Fig.~\ref{fig:T_P}(b)]
\cite{Matsukawa2016} 
and the QC
[Fig.~\ref{fig:T_P}(a)]
\cite{Deguchi,Deguchi_p}. Namely, the theory of CVF explains the fact that essentially the same quantum criticality appears in the QC and the pressure-tuned AC because the origin of CVF is the local 
CT 
fluctuation between the nearest neighbor Yb-Al sites. Since the infinite limit of the unit-cell size of the AC corresponds to the QC and the local environment around the Yb site is common in both the AC and QC, we can discuss the critical property of the QC on the basis of the theory for the pressure-tuned AC 
with the periodic lattice. 

The experimental data in the $\chi^{-1}$ vs. $T^{0.5}$ plot at low temperatures in the QC show the straight lines from the origin with almost the same slopes from $P=0$ to $P=1.6$~GPa~\cite{Deguchi,Deguchi_p}. 
This indicates that 
%
\begin{eqnarray}
\left(\frac{\partial\chi^{-1}}{\partial{P}}\right)_{T}=0
\end{eqnarray}
%
holds for low $T$ giving rise to  
%
\begin{eqnarray}
\left(\frac{\partial y}{\partial{P}}\right)_{T}=0
\label{eq:dydP_zero}
\end{eqnarray}
%
since the relation $y\propto\chi_{\rm v}^{-1}\propto\chi^{-1}$ holds near the valence QCP~\cite{WM2010}. 
This is also natural consequence of the theory of CVF for the QC that has shown the condensation of the valence QCPs in the ground-state phase diagram~\cite{WM2013}. 
If we focus on the vicinity of each one of the condensed QCPs, the concentration is 
shown
to be reflected in $\left(\frac{\partial y}{\partial P}\right)_{T=0}=0$ in $\alpha_b$. 
Namely, the ground state is singular as $1/y\to\infty$ for $T\to 0$ while the pressure derivative of $y$ becomes zero in the bulk limit of the QC. 
Then, $\alpha$ is expressed by only $\alpha_{\rm a}$ in the QC Yb$_{15}$Al$_{34}$Au$_{51}$.

The Gr\"{u}neisen parameter in the QC is obtained 
by substituting $\alpha=\alpha_{\rm a}$ and $C_{V}=C_{\rm a}-C_{\rm b}$ into Eq.~(\ref{eq:Grn}) as 
\begin{eqnarray}
\Gamma
=
\frac{C_{\rm a}}{C_{\rm a}-C_{\rm b}}\frac{1}{\kappa_T}\frac{1}{T_0}\left(\frac{\partial{T_0}}{\partial{P}}\right)_T. 
\label{eq:GrnQC}
\end{eqnarray}

To discuss the low-$T$ properties of $C_V(T)$, $\alpha(T)$, and $\Gamma(T)$ in the QC, we solve the valence SCR equation [Eq.~(\ref{eq:SCReq})] at the QCP. As the input parameters, we employ $y_0=0$ and $y_1=0.195$ which were evaluated on the basis of the periodic Anderson model in the AC for the pressure-tuned valence QCP~\cite{WM2016}. 

\begin{figure}
\includegraphics[width=7.5cm]{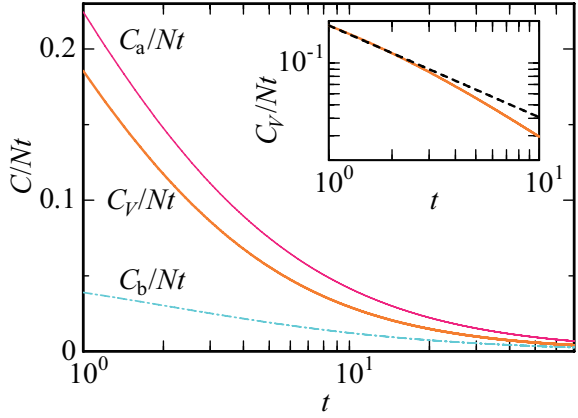}
\caption{(color online) Specific heat coefficient vs. scaled temperature $t=T/T_0$. 
$C_{V}/Nt$ 
(bold solid line), $C_{\rm a}/Nt$ (thin solid line), $C_{\rm b}/Nt$ (dash dotted line) 
at the valence QCP. Dashed line is a fit by ${\rm const}.\times t^{-0.66}$. 
}
\label{fig:Ct_t}
\end{figure}

By solving Eq.~(\ref{eq:SCReq}) as a function of $t$, 
the specific-heat coefficient $C_V/Nt$ is obtained in Eq.~(\ref{eq:Cv}) with $\left(\frac{\partial y}{\partial t}\right)_V$ being calculated, 
 which is shown in Fig.~\ref{fig:Ct_t}. 
Here, we plot the $t=T/T_{0}\gsim 1$ regime, where $\chi\propto y^{-1}\sim T^{-0.5}$ and the $T/B$ scaling were shown to appear~\cite{WM2016}. 
Note that $T$ is in the low $T$ region even if $t\gsim 1$ because $T_0$ is now extremely small 
reflecting the smallness of $A$ in Eq.~(\ref{eq:T0}), 
 which 
was explicitly shown in the periodic Anderson model in the AC~\cite{WM2016}, reflecting the locality of the (4f-3p) 
CT 
fluctuation.
As $t$ decreases, $C_{a}$ becomes dominant in $C_V$, which shows the logarithmic behavior $C_{a}/T\sim -\ln{T}$ in Fig.~\ref{fig:Ct_t}. Hence, $C_{V}/t$ at the 
lower $t$ region starts to show logarithmic increase toward $t\ll 1$
as observed in the QC~\cite{Deguchi}. 
On the other hand, the data in the region $1<t\lsim 10$ can be fit with a power-law form, giving $C_{V}/t\sim t^{-0.66}$ as shown in the inset of Fig.~\ref{fig:Ct_t}, which is also favorably compared with the experimental report in Ref.~\cite{Watanuki}.

Next, we calculate the thermal-expansion coefficient $\alpha=\alpha_{\rm a}$ by Eq.~(\ref{eq:a_a}). 
Here we input $\frac{1}{T_0}\left(\frac{\partial T_0}{\partial P}\right)_{T}=-0.44$~GPa$^{-1}$ so as to reproduce the measured lowest-temperature value of $\Gamma=-55$ in the QC (see Fig.~\ref{fig:Grun_t} below). 
The result is shown in Fig.~\ref{fig:at_t}.  As $t$ decreases, $-\alpha/t$ divergingly increases.
Here, to plot $\alpha$ in the unit of K$^{-1}$, we restored the Boltzmann constant $k_{\rm B}$ in Eq.~(\ref{eq:a_a}) and used the lattice constant $a=14.5$~\AA \ of the bcc lattice of the Tsai-type cluster in the AC. The low-$t$ part of $-\alpha/t$ is fitted with $-\alpha/t\sim t^{-0.60}$ as shown in the inset in Fig.~\ref{fig:at_t}, which is slightly different from $C_V/t\sim t^{-0.66}$ for the low-$t$ part. This is due to the contribution from $C_{\rm b}$ in $C_V$, but not $\alpha_{\rm b}$ in $\alpha$ in the QC.

\begin{figure}
\includegraphics[width=7.5cm]{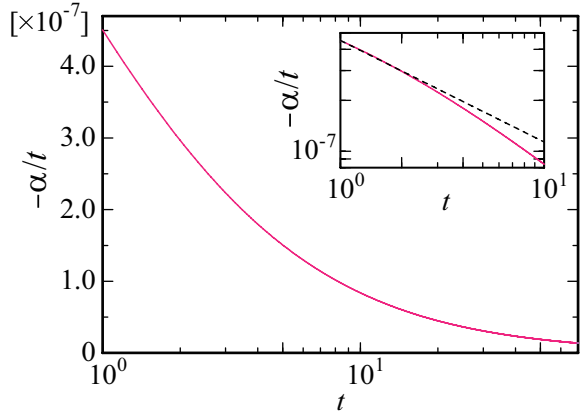}
\caption{(color online) Thermal expansion coefficient 
$\alpha/t$ vs. scaled temperature $t=T/T_0$ at the valence QCP for the QC.
Inset shows log-log plot of $\alpha/t$ vs. $t$. Dashed line is a fit by ${\rm const}.\times t^{-0.60}$. 
}
\label{fig:at_t}
\end{figure}

Then, we calculate the Gr\"{u}neisen parameter by Eq.~(\ref{eq:GrnQC}). 
Since any indication of the phase separation has not been observed in the QC, here we use 
the isothermal compressibility observed at room temperature $\kappa_T=9.6\times 10^{-3}$~GPa$^{-1}$ in the QC~\cite{Watanuki_p}. 
The resultant $\Gamma(t)$ is shown in Fig.~\ref{fig:Grun_t}. 
As $t$ decreases, $|\Gamma(t)|$ decreases as observed in the QC. 
This decrease reflects the decrease in the ratio of the specific heat $C_{\rm a}/(C_{\rm a}-C_{\rm b})$ in Eq.~(\ref{eq:GrnQC}) since $\frac{1}{\kappa_{T}}\frac{1}{T_0}\left(\frac{\partial T_0}{\partial P}\right)_T$ has no temperature dependence. 
This result gives a qualitative explanation for why $|\Gamma(T)|$ decreases on cooling in the QC.

\begin{figure}
\includegraphics[width=7.5cm]{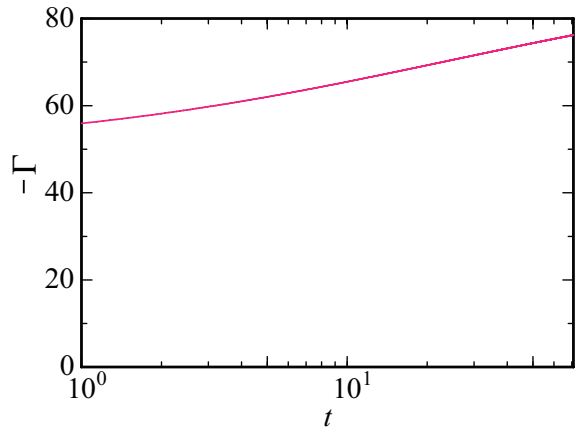}
\caption{(color online) Gr\"{u}neisen parameter  
$\Gamma$ vs. scaled temperature $t$ at the valence QCP for the QC. 
}
\label{fig:Grun_t}
\end{figure}

Non-divergent Gr\"{u}neisen parameter is ascribed to the condensation of valence QCPs in the QC. Emergence of many spots 
of the QCPs and their condensation in the ground-state phase diagram of the QC~\cite{WM2013} make the critical term in $\Gamma$ giving the divergence vanish. 

\section{Comparison with approximant crystal}
Next we discuss the meaning of the difference between the Gr\"{u}neisen parameters in the QC and AC at ambient pressure.
To analyze the Gr\"{u}neisen parameter in the QC in Eq.~(\ref{eq:GrnQC}), we derive the microscopic expression of the characteristic temperature of CVF on the basis of the periodic Anderson model for 4f and 3p electrons as 
\begin{eqnarray}
T_0=\frac{16}{3\pi^{2}
\bar{n}^{1/3}\bar{n}_{\rm c}^{1/3}}
\frac{\tilde{A}}{\bar{r}^2}T_{\rm K}, 
\label{eq:T0_TK}
\end{eqnarray}
where the Kondo temperature $T_{\rm K}$ in lattice systems is defined as
$T_{\rm K}\equiv \bar{z}V_{\rm fc}^2N_{\rm cF}$ with $\bar{z}$ being renormalization factor of quasiparticles and $N_{\rm cF}$ being density of states of conduction electrons at the Fermi level. 
Here, we have assumed that the 3p electron band is the free-electron band and the Fermi surface is spherical. 
In Eq.~(\ref{eq:T0_TK}), $\tilde{A}$ is the dimensionless coefficient in the most dominant 
CT 
(4f-3p) fluctuation defined as 
%
\begin{eqnarray}
\chi_{ii\xi\xi}^{\rm ffcc}({\bm q},0)\approx\chi_{ii\xi\xi}^{\rm ffcc}({\bm 0},0)\left[1-\tilde{A}\left(\frac{q}{k_{\rm F}}\right)^2\right]
\end{eqnarray}
%
with $k_{\rm F}$ being the Fermi wavenumber. 
Here, the irreducible susceptibility $\chi_{ii\xi\xi}^{\rm ffcc}({\bm q},i\omega_{l})$ is defined by 
%
\begin{eqnarray}
\chi_{ii\xi\xi}^{\rm ffcc}({\bm q},i\omega_{l})\equiv-\frac{T}{N}\sum_{{\bm k},m'}G^{\rm ff}_{ii}({\bm k}+{\bm q},i\varepsilon_{m'}+i\omega_{l})G_{\xi\xi}^{\rm cc}({\bm k},i\varepsilon_{m'}), 
\end{eqnarray}
%
where  $G_{jj}^{\rm bb}({\bm k},i\varepsilon_{m'})$ is the Green function for ${\rm b}={\rm f}, {\rm c}$ quasiparticle at the $j$th site with $\varepsilon_{m'}=(2m'+1)\pi{T}$ and $\omega_l=2l\pi{T}$ being the fermion and boson Matsubara frequencies, respectively~\cite{WM2016}. 
In Eq.~(\ref{eq:T0_TK}), 
$\bar{n}$
 and 
$\bar{n}_{\rm c}$
 are fillings of total and conduction electrons in the hole picture, respectively, and $\bar{r}$ is a dimensionless constant defined as $\varepsilon_{k_{\rm F}}-\bar{\varepsilon}_{\rm f}=\bar{r}D$, where $\bar{\varepsilon}_{\rm f}$ is the renormalized f level by onsite 4f-4f Coulomb repulsion, $2D$ is the width of the conduction band $\varepsilon_{k}$. 
Since the pressure dependence arises from $\tilde{A}$, $\bar{r}$, and $T_K$ in Eq.~(\ref{eq:T0_TK}), we obtain 
\begin{eqnarray}
\frac{1}{T_0}\left(\frac{\partial{T_0}}{\partial{P}}\right)
=\frac{1}{\tilde{A}}\left(\frac{\partial{\tilde{A}}}{\partial{P}}\right)
-\frac{2}{\bar{r}}\left(\frac{\partial{\bar{r}}}{\partial{P}}\right)
+\frac{1}{T_{\rm K}}\left(\frac{\partial{T_{\rm K}}}{\partial{P}}\right). 
\label{eq:T0_P}
\end{eqnarray}
By substituting Eq.~(\ref{eq:T0_P}) to Eq.~(\ref{eq:GrnQC}), we obtain 
\begin{eqnarray}
\Gamma=-\frac{C_{ a}}{C_{ a}-C_{ b}}\left[
\frac{V}{\tilde{A}}\left(\frac{\partial\tilde{A}}{\partial{V}}\right)_T
-\frac{2V}{\bar{r}}\left(\frac{\partial\bar{r}}{\partial{V}}\right)_T
+\frac{V}{T_{\rm K}}\left(\frac{\partial{T_{\rm K}}}{\partial{V}}\right)_T
\right].
\label{eq:Grn_V}
\end{eqnarray}
%

On the other hand, the AC at ambient pressure is located in the heavy-electron Fermi-liquid regime characterized by a small $T_{\rm K}$~\cite{Deguchi,Watanuki}. 
The Gr\"{u}neisen parameter 
is expressed as~\cite{Takke1981,Thalmeier1986,Goltsev2005}
\begin{eqnarray}
\Gamma_{\rm FL}=-\frac{V}{T_{\rm K}}\left(\frac{\partial{T_{\rm K}}}{\partial{V}}\right)_{S}. 
\label{eq:Grn_e}
\end{eqnarray}
For strong 4f-4f Coulomb repulsion, the Kondo temperature is expressed as 
$T_{\rm K}=D\exp(-\frac{1}{JN_{\rm cF}})$ where $J$ is the 4f-3p exchange coupling in lattice systems~\cite{Ono1989,KLM1996}. 
Then, Eq.~(\ref{eq:Grn_e}) is expressed as 
\begin{eqnarray}
\Gamma_{\rm FL}\approx-\frac{1}{JN_{\rm cF}}\left\{
\frac{V}{J}\left(\frac{\partial{J}}{\partial{V}}\right)_{S}+c
\right\},
\label{eq:Grn_TK}
\end{eqnarray}
where $c$ is a constant of $O(1)$ 
(e.g., $c=2/3$ for 
free conduction electrons). 
Since $1/(JN_{\rm cF})\approx O(10)$ holds in heavy-electron systems, $|\Gamma_{\rm FL}|$ is enhanced by this factor.  
Hence, the large Gr\"{u}neisen parameter observed in the AC at $P=0$, $\Gamma\approx -130$, is understood from Eq.~(\ref{eq:Grn_TK}). 

Since Eq.~(\ref{eq:Grn_e}) is included in the r.h.s. of Eq.~(\ref{eq:Grn_V}), the Gr\"{u}neisen parameter in the QC is also enhanced. By comparing Eq.~(\ref{eq:GrnQC}) and Eq.~(\ref{eq:Grn_e}), it turns out that the difference between $\Gamma$s in the QC $(\Gamma\approx -55)$ and AC for the measured lowest temperature at $P=0$ reflects the  volume derivative of characteristic temperatures $T_0$ and $T_{\rm K}$, respectively. 
The microscopic origin of the difference is attributed to the volume derivative of the $q^2$ coefficient of the 4f-3p 
CT 
mode 
$\tilde{A}$ 
and the relative position of the renormalized 4f level to the conduction band 
$\bar{r}$ 
from the viewpoint of the AC, which leads to the QC for the infinite limit of the unit-cell size, as seen in the 1st and 2nd terms in the r.h.s. of Eq.~(\ref{eq:Grn_V}), respectively.

\section{Theoretical predictions to experiments}
By applying pressure to the AC, the same quantum criticality as the QC $\chi\sim T^{-0.5}$ was observed at $P=P_{\rm c}=1.96$~GPa~\cite{Matsukawa2016}. 
Since the $\chi^{-1}(T)$ vs. $T^{0.5}$ plot changes its slope and intercept as $P$ varies and the intercept becomes zero only at $P=P_{\rm c}$, the AC is sensitive to pressure in sharp contrast to the QC. 
This implies that $\left(\frac{\partial\chi^{-1}}{\partial P}\right)_T\ne 0$ and hence $\left(\frac{\partial y}{\partial P}\right)_T\ne 0$ in the AC. 
Then, $\alpha_{\rm b}$ in Eq.~(\ref{eq:a_b}) contributes to $\alpha$ in addition to $\alpha_{\rm a}$, which also contibutes to $\Gamma$ in Eq.~(\ref{eq:Grn}), making 
$|\Gamma(t)|$ increase as $t$ decreases. 
Therefore, $|\alpha(T)|$ and $|\Gamma(T)|$ in the AC at $P=P_{\rm c}$ are expected to be larger than those observed in the QC. 
On the other hand, the specific heat $C_V$ is contributed from $C_{\rm b}$ as in Eq.~(\ref{eq:Cv}) in both the QC and AC. Hence, the specific heat in the AC at $P$=$P_{\rm c}$ is expected to show the similar temperature dependence to that in the QC as shown in Fig.~\ref{fig:Ct_t}. 

Furthermore, our theory is useful to classify robust criticality under pressure.  
If non-divergent $|\Gamma(T)|$ on cooling is observed in $\beta$-YbAlB$_4$ where robustness of the quantum criticality under pressure was reported~\cite{Tomita2015}, it indicates no contributions from $\alpha_{\rm b}$. This implies existence of a quantum-critical {\it phase} as in the QC, which is theoretically shown to be realized as the condensation of the valence QCPs~\cite{WM2013}.  
If diverging $|\Gamma|$ is observed on cooling, it indicates the presence of $\alpha_{\rm b}$. 
In this case there are two possibilities: One is that the QCP is located at $P=0$ but the quantum-critical {\it phase} is not realized under pressure. The other possibility is that the quantum-critical {\it phase} is realized with non-zero $\left(\frac{\partial\chi^{-1}}{\partial P}\right)_{T}$
so that the origin of the criticality robust against pressure may be different from the case of the QC.  
Hence, it is interesting to observe $\Gamma(T)$ as well as $\left(\frac{\partial\chi^{-1}}{\partial P}\right)_{T}$ in $\beta$-YbAlB$_4$ and $\alpha$-YbAl$_{1-x}$Fe$_x$B$_4$ $(x=0.014)$ in the future experiments. 


\section{Conclusion}
We have constructed the theoretical framework for calculating the specific heat $C_V(T)$, the thermal-expansion coefficient $\alpha(T)$, and the Gr{\"u}neisen parameter $\Gamma(T)$ near the valence QCP in the periodic lattice systems in the thermodynamically-consistent way. 

Then, by applying the formalism to the AC, we have discussed the specific heat, the thermal-expansion coefficient, and the Gr{\"u}neisen parameter in the QC. 
Our results explain the measured $C_V(T)$, $\alpha(T)$, and $\Gamma(T)$ at low $T$ in the QC. Since the input parameters for solving the valence SCR equation used in Sect.~3 are exactly the same as those used in Ref.~\cite{WM2016}, our results give a unified explanation for the measured magnetic susceptibility as $\chi(T)\sim T^{-0.5}$ for the zero-field limit and the $T/B$ scaling behavior in the QC as well.  
We discussed that {\it non}-divergent Gr\"{u}neisen parameter in the quantum critical QC is understood as a natural reflection of its robustness under pressure shown in Fig.~\ref{fig:T_P}(a). 
The difference in the Gr{\"u}neisen paraneters of the QC and AC is shown to reflect difference in the pressure derivative of characteristic energy scales, the CVF $T_0$ and the Kondo temperature $T_{\rm K}$, respectively. 

We have also discussed that measuring $\Gamma(T)$ provides a clear guideline to classify the origin of the robust criticality against pressure, i.e., the classification of quantum critical {\it phase}. We have made theoretical predictions for future measurements in the AC under pressure and $\beta$-YbAlB$_4$ and $\alpha$-YbAl$_{0.86}$Fe$_{0.14}$B$_4$.

Our results on the robust criticality against pressure can 
be generally applicable to other systems where the quantum critical line on the pressure axis is realized as the consequence of the condensation of many QCPs. The physical quantity expressed as the pressure derivative of the inverse valence susceptibility $y$, i.e., criticality, as Eq.~(\ref{eq:a_b}) will vanish as $\alpha_{\rm b}$ and $\Gamma_{\rm b}$ in the QC. 

\section{Acknowledgments}
We thank S. Matsukawa, K. Deguchi, N. K. Sato, and T. Watanuki for showing us experimental data prior to publications. 
This work was supported by JSPS KAKENHI Grant Numbers JP18K03542, JP18H04326, and JP17K05555.




\end{document}